# Exceptionally Fast Water Desalination at Complete Salt Rejection by Pristine Graphyne Monolayers

*Minmin Xue[*], Hu Qiu[*] and Wanlin Guo[†]*

Key Laboratory for Intelligent Nano Materials and Devices of the Ministry of Education, State Key Laboratory of Mechanics and Control of Mechanical Structures, and Institute of Nanoscience, Nanjing University of Aeronautics and Astronautics, Nanjing 210016, China

**ABSTRACT:** Desalination that produces clean freshwater from seawater holds the promise to solve the global water shortage for drinking, agriculture and industry. However, conventional desalination technologies such as reverse osmosis and thermal distillation involve large amounts of energy consumption, and the semipermeable membranes widely used in reverse osmosis face the challenge to provide a high throughput at high salt rejection. Here we find by comprehensive molecular dynamics simulations and first principles modeling that, pristine graphyne, one of the graphene like one-atom-thick carbon allotropes, can achieve 100% rejection of nearly all ions in seawater including $Na^+$, $Cl^-$, $Mg^{2+}$, $K^+$ and $Ca^{2+}$, at an exceptionally high water permeability about two orders of magnitude higher than those for commercial state-of-the-art reverse osmosis membranes at a salt rejection of ~98.5%. This complete ion rejection by graphyne, independent of the salt concentration and the operating pressure, is revealed to be originated from the significantly higher energy barriers for ions than that for water. This intrinsic specialty of graphyne should provide a new possibility for the efforts to alleviate the global shortage of freshwater and other environmental problems.

[*] These authors contributed equally to this work.
[†] To whom correspondence should be addressed. E-mail: wlguo@nuaa.edu.cn,
Tel: +86 25 84895827, 325 Mailbox, 29 Yudao street, Nanjing, 210016, China

Water is of fundamental importance not only to all living organisms including human beings, but also to most aspects of agriculture and industry such as irrigation, washing, diluting and fabricating. However, the rapid population growth and environmental pollution make clean freshwater in short supply, and a promising solution to this challenge is to desalt the saline water which accounts for 96.5% of Earth's water.[1, 2] Two approaches are commonly used in conventional desalination plants: thermal distillation, and reverse osmosis (RO) that pushes saline water of high solute concentration across a semipermeable membrane to an area of low concentration.[3, 4] Both the two technologies consume enormous amounts of energy and the output of freshwater is very limited due to the low water permeability across conventional RO membranes.[5] For instance, a state-of-the-art polymeric RO membrane that operates *via* the solution diffusion mechanism can produce freshwater at ~0.05 L per $cm^2 \cdot day \cdot MPa$ with ion rejection over 99%, and a typical RO plant must use at least 1.5 kWh of electricity to turn out 1 ton of water.[1, 2]

The high cost and low efficiency of conventional RO desalination impede the widespread use of desalination to alleviate the shortage of freshwater. To solve this issue, novel porous nanomaterials were proposed as a replacement of RO membranes in order to improve the desalination performance, as water can transport rapidly *via* their well-defined nanometer-sized pores (nanopores). For instance, hollow nanostructures including carbon[6-9] and boron nitride[10] nanotubes and nanopores drilled in one-atom-thick graphene sheets[11-13] were found to be able to transport water at least two orders faster than that predicted from the classic continuum fluid dynamics, rendering them as competent candidates for high efficient RO membranes. However, to successfully reject all ions in seawater using these nanomaterials requires the diameters of the nanotubes or graphene nanopores to be precisely controlled, the opening of the nanotubes or the edges of graphene nanopores to be chemically modified by specific functional groups, which is difficult to be guaranteed by current technology.[8, 11, 12, 14] Therefore, the exploration of new membrane materials with intrinsic well-defined nanopores of suitable size for high efficient desalination is urgently required.

Another one-atom-thick carbon allotrope that resembles graphene, graphyne, should be a promising alternative for fast and high-salt-rejection desalination. Graphyne can be regarded as formed by replacing some carbon-carbon $sp^2$ bonds in graphene with acetylenic (*i.e.*, single- and triple-bond) linkages (see Fig. 1a for graphyne geometries). Graphyne has many allotropes which differs from each other in the arrangement of carbon-carbon bonds,[15] such as α-graphyne, β-graphyne, γ-graphyne and its analogues (Fig. 1a). For instance, α-graphyne has a similar geometry as that of graphene, but with the carbon-carbon $sp^2$ bonds of graphene being replaced by acetylenic linkages. Although the introduction of the acetylenic linkages reduces somewhat the mechanical properties of graphynes, they still present a high Young's modulus in the order of 100 GPa.[16] Graphynes were demonstrated to exhibit excellent structural, mechanical, and electronic properties,[17, 18] which could be useful for applications ranging from energy storage to nanoelectronics.[19] In addition, it is surprising to note that the acetylenic linkages between two neighboring carbon hexagons form a huge number of intrinsic nanopores in graphynes (see Fig. 1a), with a surface porosity of 18%~25% comparable to that of an idealized highly-aligned nanotube array.[8] Here, we show by extensive molecular dynamics (MD) simulations that a pristine monolayer of α-graphyne, β-graphyne or graphyne-3 can desalt seawater at 100% ion rejection and produces clean freshwater at a throughput about two orders of magnitude higher than conventional RO membranes, showing a great potential to alleviate the global thirst.

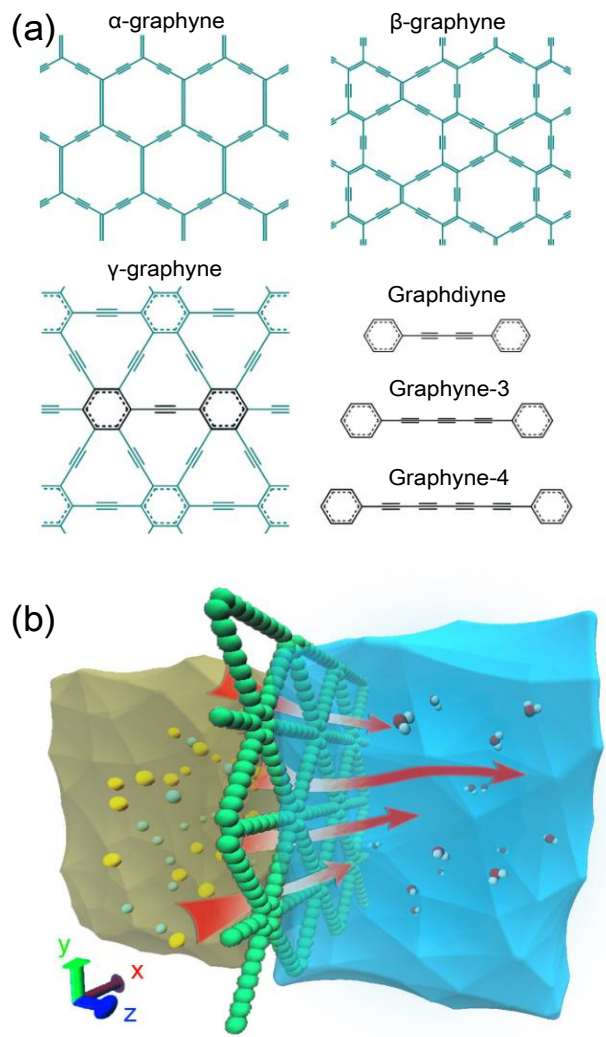

**Figure 1.** Simulation system for seawater desalination with pristine graphynes. (a) Schematic of the α-, β- and γ-graphyne sheets used in our MD simulations. The bottom right panel (from top to bottom) shows the longer acetylenic linkages of graphdiyne, graphyne-3 and graphyne-4 which share a similar geometry to γ-graphyne. (b) Schematic of the computational system. Driven by an external pressure, water molecules are allowed to pass through the intrinsic nanopores in a graphyne monolayer while ions are rejected.

## Results and discussion

The five tested graphyne sheets in the present work are α-graphyne, β-graphyne, and three analogues of γ-graphyne which can be named in terms of the number of triple-bonds between two neighboring carbon hexagons, *e.g.*, graphdiyne with two triple-bonds, graphyne-3 with three triple-bonds and graphyne-4 with four triple-bonds (Fig. 1a). The simulation system presented in Fig. 1b consists of a graphyne-3 sheet solvated in a reservoir of ionic solutions. Many members in the graphyne family including those used in the present work were predicted to be chemically stable.[20] Indeed, the successful synthesization of some substructures of graphyne has been widely reported since 1960s,[21-24] and recently it is exciting that large-area two-dimensional sheets of graphdiyne have been fabricated successfully *via* cross-linking reaction on copper surfaces,[25] paving the way for separation applications such as selective hydrogen purification.[26-28] It can be expected that the constantly emerging excellent properties of graphyne such as the unique electronic structure and the intriguing desalination performance as demonstrated in this work can stimulate experimentalists to synthesize extended graphyne materials.

**Fast water conduction across graphyne.** The most important standard determining the suitability of a membrane for desalination is the water conduction rate at high ion rejection upon the external hydrostatic pressure. We first measure the water conduction across the five graphynes under applied pressures ranging from 0 to 250 MPa. Although most industrial desalination processes operate at a pressure of 5.5 MPa, the remarkably higher pressure used here allows us to obtain reasonable statistics in a shorter simulation and the results at the lower industrial operating pressure can be easily derived. For similar reasons, $Na^+$ and $Cl^-$ ions were added to the simulation system to obtain a salt concentration of 1.2 M, which is approximately twice the average salinity of seawater. The water flux across the membrane is presented in Fig. 2 as a function of the external hydrostatic pressure. We define the net water flux as the difference between the number of water molecules per nanosecond per nanopore permeating rightward through the graphyne sheet and that permeating leftward through the graphyne. It is found that the water flux increases linearly with the pressure, consistent well with previous observations in nanotube membranes[8] and graphene nanopores.[11] Specifically, under a given pressure at 250 MPa, 5594 and 4764 water molecules were found to move rightward and leftward across the 30 nanopores of the α-graphyne membrane during the last 8 ns simulation, respectively, yielding a net water flux of ~3.2 ns$^{-1}$ per nanopore (see Fig. 2 and Table 1). Despite sharing similar pore geometry, β-graphyne can transport water about 31% faster than α-graphyne, resulted from its slightly larger effective internal pore radius (see Table 1).

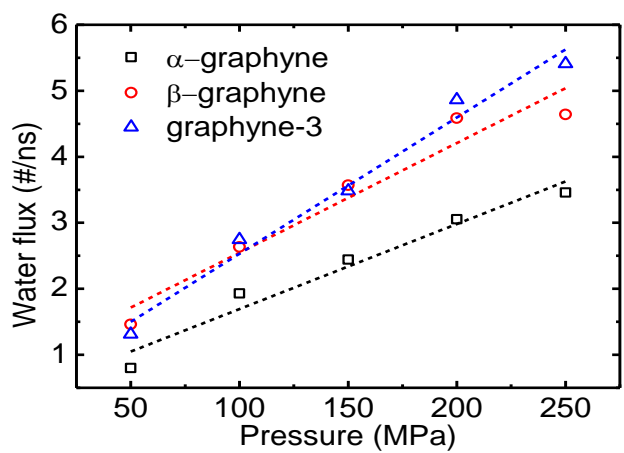

**Figure 2.** Water flux per nanopore across α-graphyne，β-graphyne and graphyne-3 as a function of the external pressure. The dashed lines are linear fitting to the data.

**Table 1.** Desalination performance of graphynes at 250 MPa (Units: $ns^{-1}$ per nanopore).

| Type | Effective internal pore radius (Å) | $H_2O$ | $Na^+$ | $Cl^-$ | $K^+$ | $Mg^{2+}$ | $Ca^{2+}$ |
|---|---|---|---|---|---|---|---|
| α-Graphyne | 1.746 | 3.2 | 0 | 0 | 0 | 0 | 0 |
| β-Graphyne | 1.773 | 4.2 | 0 | 0 | 0 | 0 | 0 |
| Graphdiyne | 1.002 | 0 | 0 | 0 | 0 | 0 | 0 |
| Graphyne-3 | 1.736 | 5.6 | 0 | 0 | 0 | 0 | 0 |
| Graphyne-4 | 2.470 | 10.0 | 0.026 | 0 | 0.005 | 0 | 0 |

As an analogue to γ-graphyne, graphdiyne was found to be impermeable to water or ion during our simulations at a hydrostatic pressure up to 250 MPa (Table 1), suggesting that such a graphyne with small intrinsic nanopores cannot serve as a desalination device. The remaining two γ-graphyne analogues tested in the present work, graphyne-3 and graphyne-4, are both permeable to water and present higher water permeabilities than those for α-graphyne and β-graphyne (see Table1). It is noteworthy that the effective internal pore radius of graphyne-3 or graphyne-4 is indeed smaller compared to that for α-graphyne or β-graphyne, thus the high water permeabilities should be attributed to the significantly larger open pore area of the intrinsic triangular pore of graphyne-3 or graphyne-4. The porosity is an important measurement of the desalination performance of a membrane, and can be obtained by calculating the percentage of the open pore area in a membrane material. For instance, the effective open area of an α-graphyne nanopore is ~10.6 $Å^2$ assuming the van der Walls radius of carbon to be 1.7 Å, leading to a surface porosity as high as 25.7%. With this high porosity, the throughput of an ideal α-graphyne membrane is estimated to be ~8.1 L per $cm^2$ day MPa if used in industrial RO desalination. It is surprising to find that the final throughput of β-graphyne (5.5 per $cm^2$ day MPa) is much lower than that of α-graphyne due to its evidently smaller porosity, athough β-graphyne presents a faster water flux per each intrinsic nanopore (Table 1). Similarly, the throughputs of graphyne-3 and graphyne-4 are 11.7 and 14.3 L per $cm^2$ day MPa, respectively. Noting that the fast water permeability demonstrated here is achieved using a pristine graphyne without any postprocessing such as chemical modifications, showing the feasibility and convenience of the graphynes as efficient RO membrane materials.

**Water structure in the conduction pore of graphyne.** Water molecules are found to form a single-file structure in the conduction pore of graphyne during our simulations except that of graphyne-4 (see inset of Fig. 3b). The water oxygen density profiles along the direction normal to the graphyne sheets (*i.e.*, +*z*) were obtained by measuring the average number of water molecules within a 2-Å thick slab during the last 8 ns simulations (Fig. 3a). In general, the density profiles of all tested graphynes are similar to each other and to that of a pored graphene,[13] with two peaks near the graphyne sheet (*z*=0) showing the two binding sites. This fact indicates that the specific geometry that whether the intrinsic pore is triangular or hexagonal has little impact on the general water distribution along the conduction path. We also calculate the water oxygen density maps inside the graphyne pores to further understand the distribution character of water when transporting across the pores (shown in Fig. 3c). For hexagonal pores of α-graphyne and β-graphyne, it is unsurprising to find that the shape of the density distribution map is round, with the highest probability for water occupation at the pore center (two top panels in Fig. 3c). In the case of the triangular pores in graphyne-3 or graphyne-4, what we obtain is a triangular density distribution map (two bottom panels in Fig. 3c). It is also found that one area is available for water passage across graphyne-3 (bottom left panel in Fig. 3c), while three apparent areas for water passage are determined for graphyne-4, which are symmetric around the pore center (bottom right panel in Fig. 3c).

We then calculated the probability distribution of the dipole orientation of the single-filed water within the conduction pore of graphyne (*i.e.*, water molecules between the two peaks in the density profile shown in Fig. 3a). The dipole orientation is defined as the angle between the water dipole vector and the $z$ axis. The monitoring of the MD trajectory suggests that the dipole orientation of the single-filed water reverses frequently during our simulations, but its averaging does tend to stay at ~40 ° and ~140 ° (see Fig. 3b). The concerted hopping of such a water single-file should be essential for the exceptionally fast water passage across graphyne .

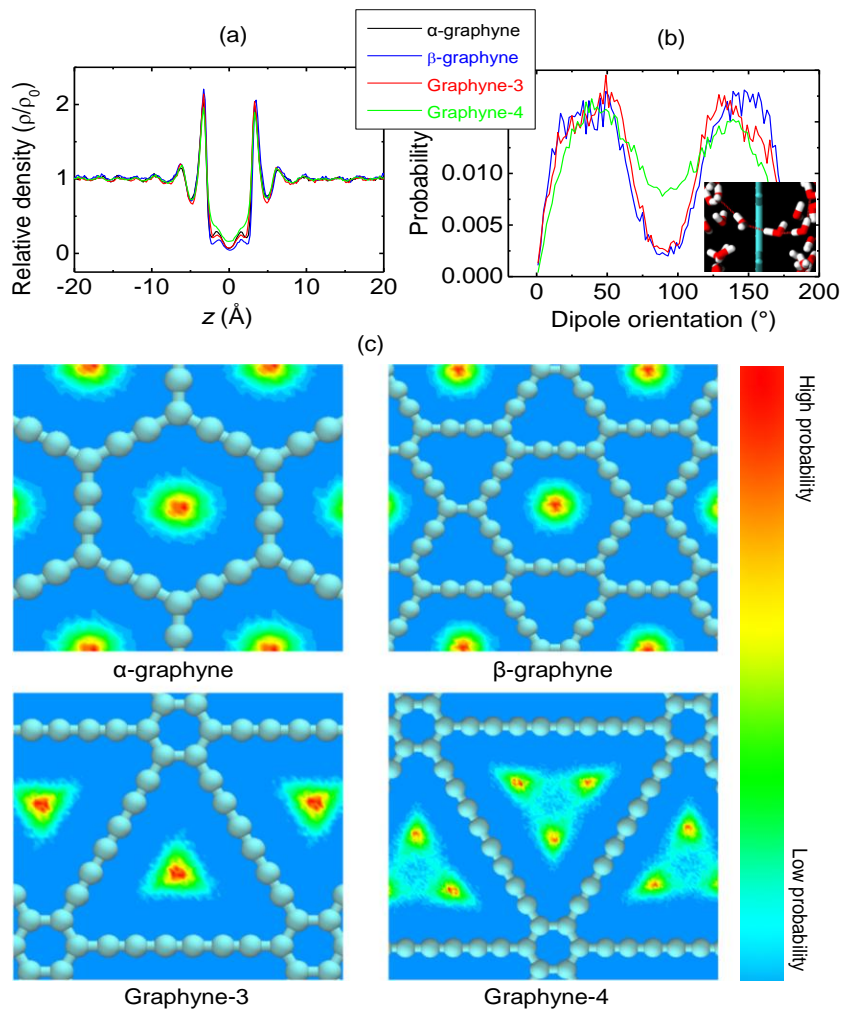

**Figure 3.** Water structural properties inside different graphyne pores. (a) Distribution of water density $\rho$ along $z$ axis relative to the bulk water density $\rho_0$. The two peaks near the graphyne sheet correspond the two binding sites for water transporting across graphyne. (b) Dipole orientation of water molecules inside the graphyne pores. The dipole orientation is defined as the angle between the water dipole vector and the *z* axis. Inset is a snapshot of single-filed water molecules. (c) Water oxygen density maps inside different graphyne pores. Light blue and red indicate regions with the lowest and highest probabilities to find an oxygen atom, respectively.

**Complete salt rejection.** Another key issue to a desalination membrane is the percentage of salt rejection. For α-graphyne, β-graphyne and graphyne-3, no $Na^+$ or $Cl^+$ ion was observed to pass through the sheets during our 10 ns simulations under a hydrostatic pressure up to 250 MPa, yielding a complete salt rejection of 100% (Table 1). Furthermore, this perfect ion rejection ability of a pristine graphyne still exists when increasing the solution concentration from 1.2 M to 3.6 M (data not shown), in sharp contrast to the concentration-dependent salt rejection of carbon nanotubes at a concentration higher than 0.01 M.[14] To explore the ability of rejection towards other ions in seawater, we carried out a MD simulation at a pressure of 250 MPa with a reservoir containing 0.6 M $Mg^{2+}$, 0.12 M $K^+$, 0.12 M $Ca^{2+}$, and 1.56 M $Cl^-$. It is interesting to find that none of these ions can permeate through α-graphyne, β-graphyne and graphyne-3, showing the robust ability of salt rejection (Table 1). We also performed a simulation with the salt water sandwiched between two graphyne-3 sheets in order to obtain a visualized figure of the industrial RO desalination process, and found that similar water permeability and 100% ion rejection can also be achieved (see Fig. S2 in Supporting Information). For graphyne-4 with relatively larger pores, $K^+$ and $Na^+$ can permeate through the membrane (see Table 1), whereas some $Cl^-$ ions were found to occasionally linger around the pore area but still cannot pass through. Examination of the MD trajectory further suggests that the two larger ions, $Mg^{2+}$ and $Ca^{2+}$, cannot even approach the pore area of graphyne-4.

To explore the underlying mechanism of the demonstrated complete ion rejection of graphyne, we calculated the PMF free energies for water molecule and ions across α-graphyne nanopores using umbrella sampling. The results in Fig. 4 show that water can pass through graphyne with an energy barrier less than 2 kcal/mol. Furthermore, the two minor valleys observed in the water free energy profile (two arrows in Fig. 4) suggest that water transport across graphyne can adopt a stepping mode by first jumping to these valleys and then climbing over the highest barrier at the graphyne plane ($z=0$). Such a stepping process makes water much easily to pass through the graphyne pores, leading to the exceptionally fast water permeability. In sharp contrast, the demonstrated energy barriers opposing the passage of $Na^+$, $K^+$ and $Cl^-$ ions are up to ~10 kcal/mol. Furthermore, no minor valleys were observed within the conduction path of these ions, and they must jump across the highest energy peak within a single step. These facts make ions difficult to pass through the graphyne pores. For the divalent cations, $Mg^{2+}$ and $Ca^{2+}$, the energy barriers opposing ion conduction are as high as ~60 kcal/mol (see Fig. 4 inset). Such high energy barriers should be caused by the sieving effect since the α-graphyne nanopore is too small to accept a divalent ion into its interior.

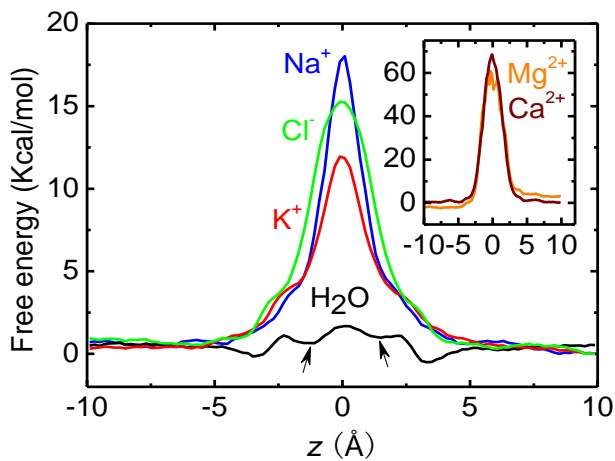

**Figure 4.** PMF free energies for water and ions across the α-graphyne monolayer.

**Performance comparison of graphynes with conventional RO menbranes.** The above systematic MD simulations have examined the performance of three graphyne membranes for seawater desalination. Graphdiyne is impermeable to water and ions due to the limited pore size. α-graphyne, β-graphyne and graphyne-3 can rapidly desalt seawater at 100% salt rejection, indicating that these pristine graphynes can act as high-performance RO desalination membranes. Graphyne-4 can conduct water more quickly than other graphynes, but at a weakened salt rejection. We compare the performance of the graphyne desalination devices with the widely used conventional RO membranes in Fig. 5. It is surprising to note that the permeability of α-graphyne, β-graphyne and graphyne-3 at 100% salt rejection can be about two orders of magnitude faster than the conventional RO membrane at ~98.5% salt rejection (*i.e.*, BWRO).

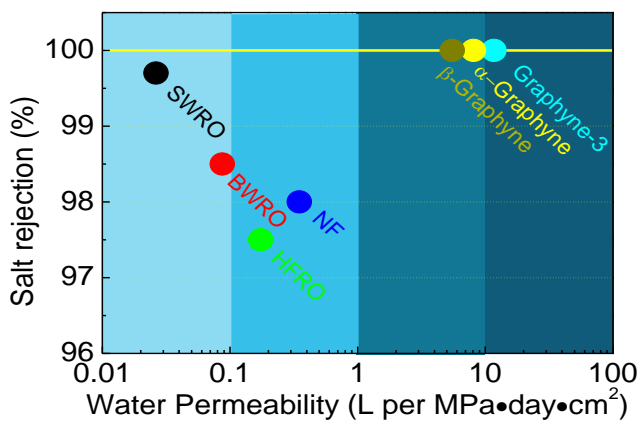

**Figure 5.** Performance comparison between α-graphyne, β-graphyne, graphyne-3 and the conventional RO desalination membranes. The three graphyne monolayers can desalt water at 100% ion rejection with a throughput significantly faster than commercial RO membranes such as polymeric seawater RO (SWRO), brackish water RO (BWRO), high-flux RO (HFRO) and nanofiltration (NF) .[3]

We have demonstrated that α-graphyne, β-graphyne and graphyne-3 can serve as high efficient desalination membranes, which holds the promise for an excellent alternative for commercial RO desalination. Compared to other nanomaterials such as carbon nanotubes and pored graphene based membranes, the pristine graphyne membrane has many apparent advantages due to its intrinsic polyporous structure. First, it is quite difficult to precisely control the pore size of pored graphene, or the diameter and alignment of carbon nanotubes in the membranes, while a pristine graphyne sheet has intrinsic well-defined nanopores of the same size. Second, the high permeability or ion rejection of graphene nanopores and carbon nanotubes rely on specific chemical functionalizations in the tube ends or pore edges,[7, 11, 12] while the excellent desalination performance can be readily realized for pristine graphynes. As a result, the intrinsic structural advantage makes α-graphyne, β-graphyne and graphyne-3 very promising materials for future water desalination.

The two key issues to the practical application of graphyne in desalination are the synthesis in large area and the mechanical strength of the monolayer material. As a novel nanostructure, the fabrication of graphyne family has attracted more and more attentions due to its excellent structural, mechanic and electronic properties,[16, 17] and recently the two-dimensional sheet of graphdiyne has been successfully fabricated. We believe that the present finding of the excellent desalination ability of graphyne can stimulate experimentalists to continue their efforts. The membrane integrity of graphyne upon external hydrostatic pressure during desalination is also an important issue. It can be expected that the mechanic strength of graphyne monolayer against the applied pressure can be enhanced by putting them on supporting porous polymeric layers that are widely used in conventional RO industry.

## Conclusions

In summary, we show by systematic MD simulations that certain pristine graphyne membranes can transport water rapidly in a single-file manner while rejecting ions at 100%, serving as an efficient desalination device. Such a complete ion rejection holds good for nearly all ion types existing in seawater such as $Na^+$, $Cl^-$, $Mg^{2+}$, $K^+$ and $Ca^{2+}$, resulted from the higher energy barriers opposing ion conduction than that for water. The demonstrated freshwater throughput of α-graphyne, β-graphyne and graphyne-3 is about two orders of magnitude faster than conventional RO membranes. Considering the intrinsic advantages of graphynes for desalination including the uniform size of conduction pores and the pristine pore edge, we believe that such a desalination membrane can provide a feasible route to significantly improve the performance of existing desalination technologies to alleviate the global thirst.

**Materials and Methods**

We first optimized the geometric structures of graphynes with density-functional theory calculations (see Supporting information for details). The yielded bond lengths and lattice parameters are comparable to those given in previous studies.[18, 20, 29] Then, a rectangular graphyne sheet shown in Fig.1a was then constructed from the relaxed structure and fixed with its center at zero during MD simulations, separating the reservoir into two regions. All MD simulations were carried out with NAMD2[30] and visualized with VMD.[31] Water was modeled using the TIP3P model[32] and the carbon atoms were treated as Lennard-Jones particles with parameters taken from those for aromatic carbons (CA) in the CHARMM27 force field.[33] A series of test simulations with the $sp^2$ carbon in AMBER force field[34] being used to describe carbon atoms yield similar desalination performance (see Fig. S3 in Supporting Information). Periodic boundary conditions were applied in all directions, and the particle mesh Ewald (PME) method[35] was employed to treat the long-range electrostatic interactions. A time step of 1 fs was used and coordinate data was collected every 1 ps. The NVT canonical ensemble (constant number of particles, volume and temperature) was adopted with the temperature at 300 K controlled by the Langevin dynamics.[36] After 1000 steps of energy minimization, the system was initially equilibrated for 0.5 ns. Then we performed MD simulations with applied hydrostatic pressures to examine the desalination performance of the five graphynes. As adopted previously,[37] the hydrostatic pressure difference $\Delta P = nf/A$ was created along +z direction by applying a constant force $f$ to the oxygen atoms of all water molecules in the region of -20 Å < $z$ < -15 Å, where $n$ is the number of water molecules with an external force and $A$ is the cross-sectional area of the system. Each simulation with hydrostatic pressure was performed for 10 ns, with the last 8 ns being used for data analysis.

Potentials of mean force (PMF) for water or ions were calculated along the reaction coordinate (+z) with umbrella sampling,[38] with the width of each umbrella window at 0.5 Å. A biasing potential of 20 kcal/mol/Å$^2$ was applied on the $z$ coordinate of water oxygen or ion in each window. All simulations were conducted for 500 ps, with the last 400 ps for data analysis. The yielded umbrella histograms were then unbiased and combined using the weighted histogram analysis method (WHAM)[39, 40] to obtain the PMF profile.

**Note added before submission**

When we finished this manuscript, we lerarned of two papers reported by independent groups that work on a same topic[41, 42]. Only one family of graphyne, graphyne-n , was considered in these studies. In constrast, we report here a systematic understanding of the high performance desalination using graphyne-n, as well as α-graphyne and β-graphyne.

**Supporting Information Available:** Fig S1-S3. This material is available free of charge *via* the Internet at http://pubs.acs.org.

**Author Contributions**

M.X. and H.Q. contributed equally.

**Notes**

The authors declare no competing financial interest.

**Acknowledgments**

This work was supported by 973 program (2013CB932604, 2012CB933403), and the National NSF (91023026) of China.

TABLE OF CONTENTS GRAPHIC

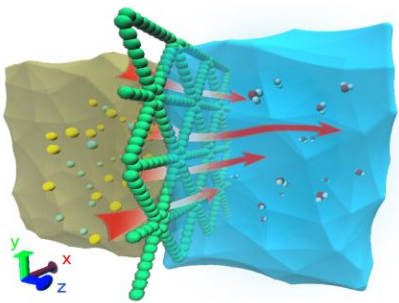